 \documentclass[12pt,preprint]{aastex}
\usepackage{graphicx}
\usepackage{natbib}

\def \be {\begin{equation}}
\def \en {\end{equation}}

\def \mt { }
\def \mttt {}

\def \mta { }

\def \mtb {  }

\def \mtbb { }
\def \mtcc {  }
\def \mtdd { }

\def \mtv {  }
\def \mtg {  }

\def \mtx {  }
\def \mtxx {  }
\def \mtxy { }
\def \mtyy { }
\def \mty { }

\def \ff{  }
\def \ww{  }

\shorttitle{...}

\shortauthors{....}

\begin{document}

\title{\Large An Emerging Class of Gamma-Ray Flares from Blazars:
Beyond One-Zone Models}

\vspace{2cm}
\author{\large
M.~Tavani\altaffilmark{1,2,3}, V.~Vittorini\altaffilmark{1},
 A.~Cavaliere\altaffilmark{1,2}\\
 \bigskip
\normalsize{ Submitted to the \textit{Astrophysical Journal}: July
23, 2015. Accepted: October 13, 2015}
 {\altaffiltext{1}{INAF/IAPS--Roma, Via del Fosso del Cavaliere
100, I-00133 Roma, Italy} \altaffiltext{2}{Univ. ``Tor Vergata'',
Via della Ricerca Scientifica 1, I-00133 Roma, Italy}
\altaffiltext{3}{Gran Sasso Science Institute, viale Francesco
Crispi 7, I-67100 L'Aquila, Italy}}}

\vskip .1in


\begin{abstract}

Blazars radiate from relativistic plasma jets with  bulk Lorentz
factors $\Gamma \sim 10$, closely aligned along our line of sight.
 In  a number  of blazars of {the} Flat Spectrum Radio
{Quasar  type}  such as 3C 454.3 and 3C 279 gamma-ray flares
{have} recently {been} detected with very high luminosity and
little or no counterparts in the optical and soft X-ray bands.
They challenge the current one-zone leptonic models of emissions
from within the broad line region. {\mtv The latter} envisage the
optical/X-ray emissions  to be produced  {as} synchrotron
radiation   {by} the same population of  { highly relativistic}
electrons  {in the jet} that  {
would also yield} the
gamma rays by inverse Compton up-scattering  {of surrounding soft
photons}. To meet the challenge we present here a model based on
primary synchrotron photons emitted in the broad line region by
 {a plasmoid moving out with the jet} and
 scattered back toward the incoming plasmoid by  {an outer} plasma
clump acting as a mirror. We consider both a scenario based on a
\textit{static} mirror located outside the BLR, and
 {\mtv an} alternative provided by a \textit{moving} mirror
geometry. We show that  mirroring phenomena can locally enhance
the  density and  {\ww anisotropy with associated}  relativistic
boosting of soft photons within the jet,  so as to trigger bright
inverse Compton gamma-ray transients  from nearly steady
optical/X-ray synchrotron {\mtv emissions}. In this {picture} we
interpret the peculiarly asymmetric lightcurves of the recently
detected gamma-ray flares from 3C 279. Our scenario provides a
promising start to {understand the widening} class of bright and
transient gamma-ray activities in blazars.

\end{abstract}


\section{Introduction}

{\mtxx Blazars are active galactic nuclei (AGNs) whose emissions
{\mtxy are }  dominated by the Doppler boosted radiation from
relativistic jets (Urry \& Padovani 1995).}
 Blazar radiations - differently from plain Quasars' - are
highly non-thermal. They are  powered by  a central supermassive
black hole (BH) that launches the jets closely along  our line of
sight  with {\mttt considerable} bulk Lorentz factors $\Gamma \sim
10$. In the jet, highly relativistic electrons with random Lorentz
factors up to $\gamma \sim 10^3$ produce the radiations we
observe.
 These bear a clear non-thermal mark in
their spectral energy distribution  (SED) that  gathers into two
main humps, see Fig. 1. An optical-UV peak extending out to soft X
rays is commonly understood in terms of synchrotron (S) emission
by the relativistic electron population in the jet magnetic field
that attains values  $ B \sim \,1\,G$ at $R \sim 3\cdot 10^{17}$
cm from the BH. A second peak in the GeV range is widely discussed
in terms of inverse Compton up-scattering by the same electron
population of soft ("seed") photons\footnote{\mtv For standard
expressions of the  powers and frequencies  radiated by the
synchrotron and inverse Compton processes we refer the reader to
Vittorini et al. 2014, Appendix. {\mtxx In the following, it will
be useful to keep in mind that the observed luminosity (for the
usual isotropic evaluation) or the corresponding SED
%
follow the proportionality $L \simeq \epsilon \, F_{\epsilon}
\propto \Gamma^4 \, U' \, V'$, in terms of the 
{\ww emitted} photon energy $\epsilon$, the energy density {\ww of
seed photons} $U'$ and the emitting volume $V'$ {\ww both} in the
jet comoving frame.}} (see, e.g., Sikora, Begelman \& Rees, 1994).

The seeds may just comprise the  contribution  by the very S
emission in the jet, {\mtg the so called synchrotron self-Compton
(SSC) process {\mtxx (e.g., Maraschi et al. 1992, Bloom \&
Marscher 1996)}. } On the other hand, in the external Compton (EC)
radiation mode {\mtxx the seeds} are dominated by the 
photons produced in the optical Broad Line Region (BLR) at $R \sim
3 \; 10^{17}$ cm, or by the dusty infrared {\mtxx torus} at
somewhat larger distances {\mtxx (e.g., Dermer et al. 1992, Sikora
et al. 1994)}.
{\mtxx Both regions} reprocess/reflect   the UV glow of the inner
accretion disk around the BH. In fact, the relative heights of the
S and of the inverse Compton peaks mark the two basic Blazar
flavors: the BL Lac type with comparable {\mtyy heights} {\mtg for
which the SSC process may be adequate,} and the Compton-dominated
(and so gamma-ray dominated) Flat Spectrum Radio Quasars (FSRQs)
{\mtxx for which} a substantially larger density of seed photons
is needed {\mtxx (e.g., Sikora et al. 2009, Boettcher et al.
2013).}

\begin{figure}
  \centerline{\includegraphics[width=15cm, angle = 0]{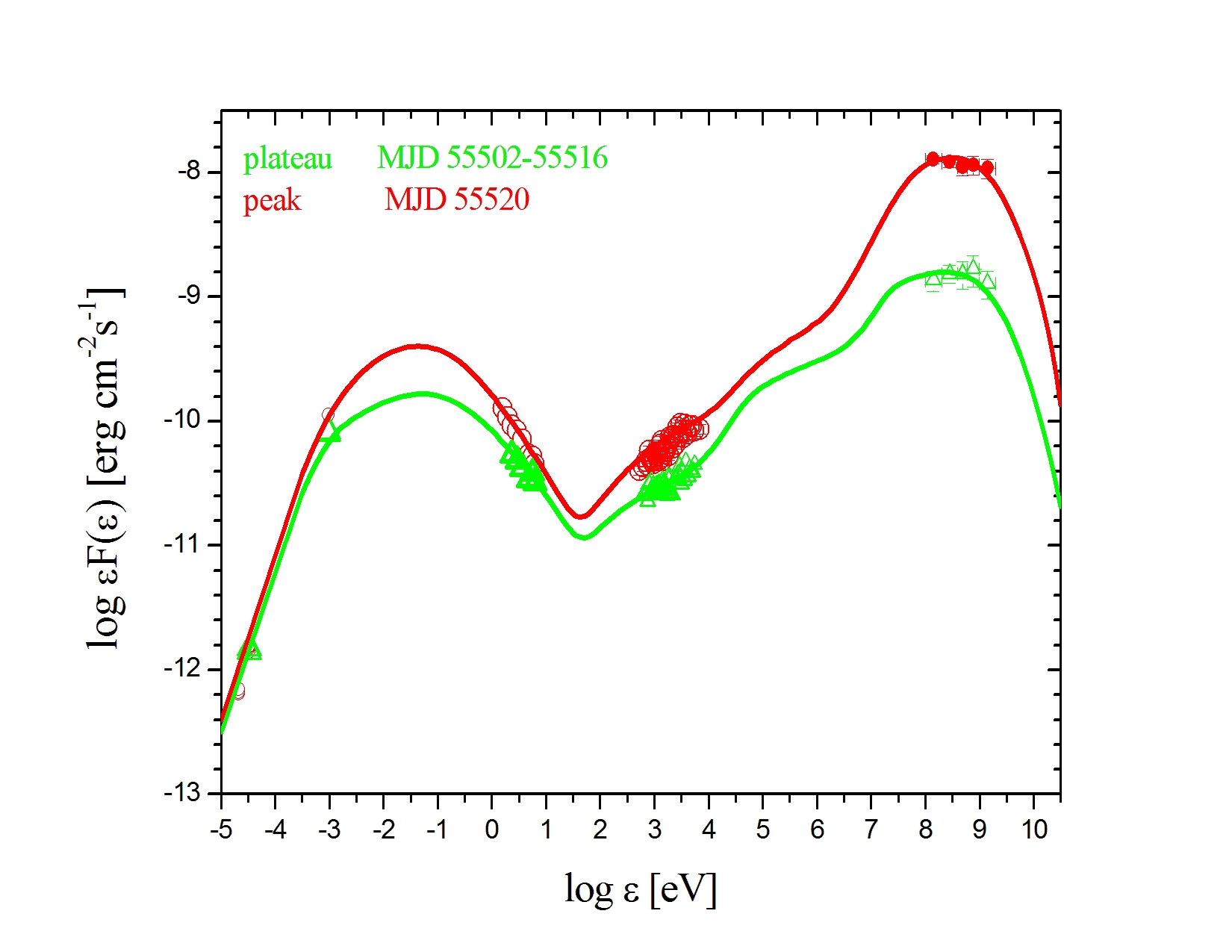}}
    \caption{The two-humped Spectral Energy Distribution from a
    FSRQ Blazar with large Compton dominance. Here it is represented
    the behavior of 3C 454.3 in November 2010, when the Compton dominance
    was particularly enhanced around the MJD 55520. {\mtv Details of the spectral
    fit represented here
    can be found in V14; 
    {\mtxx in particular,} the magnetic field
     {governing }the synchrotron emission
    is $B \simeq 1$ G, and  the high-energy electron distribution
    is modeled as a {\mtxx flat} power-law 
    {\ff with break energy} $\gamma \simeq 10^3$.}}
\label{fig-basics}
\end{figure}

Blazars of both flavors have been found to be highly variable, and
particularly so in gamma rays. In fact, a new class of gamma-ray
Blazars of the FSRQ flavor is now emerging that is marked by
flares with extreme properties such as:

\begin{itemize}

\item[1.] very large gamma-ray luminosities up to $L_{\gamma} \sim
10^{48}$   erg/s {in the range 100 MeV - 10 GeV}, together with a
large ratio $L_{jet} / L_{disk} \sim 10$;

\item[2.] strong Compton dominance with  $L_{\gamma} \gg L_{S}$,
only marginally correlated in time with optical emissions and even
less with soft X rays;

\item[3.] short variability time scales, down to a few hours;

\item[4.]  unusually hard gamma-ray spectra corresponding to SEDs
flat or even rising  {beyond} 100 MeV with photon index 
{\ff as hard as}  $1.6$, that in the EC process imply   flat
electron energy distributions up to {\ff  a break value} $\gamma
\sim 3\cdot 10^3$;

\item[5.] occasional spikes  at TeV energies.

\end{itemize}

 In addition to 3C 454.3 (e.g.,Vittorini et al. 2014) and 3C 279 (Giuliani et al. 2009, Hayashida et al. 2012, 2015) several
prominent FSRQs show similar features, {\mtg at least episodically
including PKS
1510-089 (Abdo et al., 2010a, Saito et al., 2013), 
PKS 1830-211 (Donnarumma et al., 2011), OJ 248 (Carnerero et al.,
2015).}
It is becoming clear that the full  behavior of  {\mtxx blazar}
sources is complex and not  easily amenable to a single and
universal source structure. A widely entertained picture
addressing the emissions from FSRQs has been  based on the
so-called one-zone modeling, with S and EC radiations produced in
the \textit{same} region within, or close to the BLR at  $R \sim
0.1 $ pc {\mtxx (for BLR properties, see Peterson 2006)} . This
constitutes the canonical picture that {has} to withstand the
wealth of incoming {\mtxx new} data.

However, such models have been recently shown by Vittorini et al.
2014 (hereafter V14) to fail in accounting for the strong
gamma-ray flaring of 3C 454.3. A similar conclusion is 
{\mtx implied} by Hayashida et al. (2015, hereafter H15)  in the
case of the 2013-2014 gamma-ray activity of 3C 279.
 Both sources  have clearly shown
episodes of \textit{decoupled} gamma-ray and optical/X-ray
radiations, with  complex and specific timings which include:

\begin{itemize}

\item stretches of enhanced, flickering gamma-ray emission
spanning weeks or months (that we name "plateau");

\item  on top of a plateau, day-long gamma-ray flares attain
luminosities $ L_\gamma  \gg L_S$ having little if any
counterparts in the optical band and in soft X rays, whilst some
optical flashes occur with no corresponding gamma-ray counterpart;

\item   short time scales down to a few hours, with even faster
rise times marking  the truly bright flares;

\item  slow ups and downs  occurring in the optical and/or soft
X-ray bands on scales of several months to years, with  mild
maxima and  no detailed correspondence with the gamma-ray flares.

\end{itemize}

 {Note that } episodes of gamma-ray activity including sharp
flares often occur near the top of such  secular enhancements in
the optical or soft X-ray bands. {\mtb Furthermore, }  a location
of the EC source in an extended environment as rich in UV photons
as the BLR would imply a prompt degradation of the spectra  at  $h
\, \nu   > 10 $ GeV by photon-photon interactions producing
electron - positron pairs. {\mtxy This process would} prevent hard
spectra from  {reaching the observer}.

Our {\mtb paper} will address these issues and is structured as
follows. We will start with a discussion of observed time scales
and features of the gamma-ray vs. optical/X-ray radiations,
focused on the two prominent FSRQ sources 3C 454.3 and 3C 279. We
{\mtx discuss how the complex correlation patterns in these bands}
challenge the canonical one-zone structure for the sources. We
then 
{\mtx summarize  the basic idea } 
{\mtxx underlying a} viable alternative to one-zone models that we
proposed for 3C 454.3 (V14). {\mtx This is based
on a \textit{mirror-driven process} within the jet for inducing} 
{ localized} and
transient enhancements of S photon density 
\textit{beyond} the BLR. {\mtx The process naturally provides}
{\ww localized} seed photon densities {\mtx large} enough for intense 
{\ww and  
short EC } production of gamma rays. {\mtx We consider separately
static and
moving-mirror scenarios} {\ff and carry through our discussion on} 
{\mtxy how they apply} {\ff to gamma-ray flaring blazars}.
{\ff In particular, we } {\ff focus on} the flares of 3C 279 in
2013-2014
and  on their 
extreme features. {\mtyy We
finally}  discuss  
{\mtx how our physical picture   is related with}
dissipation of magnetic energy {\mtx ultimately} 
{\ff producing} both the dense photon bath and the acceleration of
highly relativistic electrons that concur to feed the EC
radiations.

\section{The challenge from multiple time scales and missing correlations}

As anticipated in Sect. 1 most leptonic  models for FSRQs are
based on electron S emission prevailing up to the UV and soft
X-ray bands, and on inverse Compton  {radiation} taking over
beyond. The latter process is fed when the relativistic electrons
in the jet up-scatter seed photons provided by internal S emission
and/or by external sources like the UV accretion disk and its
reprocessed radiations such as the optical lines and the IR torus
emissions. The broad-band SED is shaped at any moment by the
prevalence of one or the other process.

The one-zone models, in particular, envisage radiations from the
\textit{same} region, typically  the BLR at distances $R_{BLR}
\simeq 3 \cdot 10^{17}$ cm, and so predict neatly
\textit{correlated} emissions from the optical to the gamma-ray
band  over most time scales, {\ff as discussed by, e.g., Paggi et
al. (2009) for the standard SSC radiation.} However, as we show in
detail below, such a {\mtg strong}  correlation is itself
challenged {\ff by the data}, particularly by several detailed
observations of 3C 454.3 and 3C 279.
In fact, these FSRQs show a number of \textit{diverse} time scales
that  particularly differ from gamma to  X rays. For strong
gamma-ray flares the risetimes can be as short as 1 hr (limited by
the {\ff effective} instrumental time resolution); on the other
hand, the plateaus
last up to several weeks.  Meanwhile, the X rays  
{\mtxx show} just mild undulations on scales of several months
{\ff (see Fig. 2)}. Additional constraints are set by the  hard
observed spectra that require in and around the source a low
optical depth for pair producing photon-photon interactions.

Meeting all {\ff such} challenges together  apparently requires
that in different bands not only different source geometries
apply, but also different  physical processes  proceed
independently of each
other. Our 
{\mtxx aim} is to see whether  {\ff in fact} these can be
\emph{related} in a wider physical picture, and connected as
tesserae composing a comprehensive pattern.

\subsection{The case of 3C 454.3}

3C 454.3 in particular 
{\ff features} stretches of high-energy  activity lasting
months/years but also long, inactive {\ff states}.  Since the
beginning of systematic gamma-ray observations (and related
multifrequency monitoring) in 2007, 3C 454.3 displayed several
very intense gamma-ray flaring episodes, notably in November 2009
and 2010 (Vercellone et al. 2010, Striani et al. 2010, Vercellone
et al., 2011, Wehrle et al. 2012, Jorstad et al. 2013).

In particular, the November 2010 episode featured {\ff for several
days} the most intense gamma-ray flaring source ever detected, an
episode that was notably 
{\ff superimposed on}  an enhanced plateau emission lasting
several weeks. {\mtyy Such sequences (short flares on top of a
long plateau)} had been previously noted in the EGRET data, a
notable example being that of PKS 1622-287 (Mattox et al. 1996).
However, EGRET could not follow in detail the lightcurve for
extended times, and therefore such a kind of features remained
neglected.

On the other hand, AGILE and Fermi by their long pointings
provided extended lightcurves, and could test whether the sequence
"flare on plateau" constitutes an exception or rather a common
mode of strong gamma-ray flaring in FSRQs.
In particular, the very
intense and complex 
flare of November 2010 was focused and modelled by V14. In that
case, the Compton dominance  was enhanced {\mtyy in the flare}
relative to the plateau -- itself up by a factor of 2 --  by a
further factor of 6.

\begin{figure}
  \centerline{\includegraphics[width=15cm]{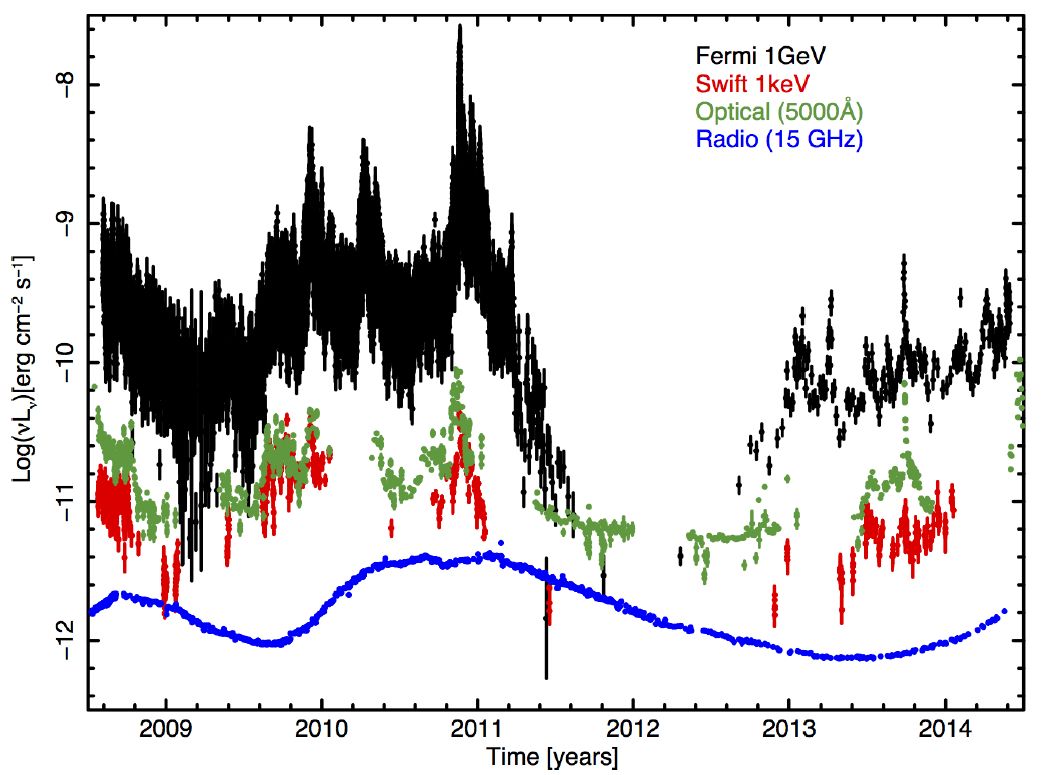} }
    \caption{Multifrequency monitoring of 3C 454.3 (from Giommi et al. 2015). }
\label{fig-3c454}
\end{figure}

\subsection{The extreme case of 3C 279}

3C 279 shows - in a context of 
flares on top of a plateau - {\mtx several episodes of enhanced}
gamma-ray emission with even {\mtx larger} Compton dominance,
extended spectral hardness, and lack of optical/X-ray
correlations. {\mtx Since 2007, 3C 279 has been monitored above
100 MeV by AGILE (Giuliani et al. 2009) and Fermi (Abdo et al.
2010b, Hayashida et al. 2012, 2015).}
Its recent activity during the period 2013-2014 was 
{\mt reported} by H15 {\mtx (see  Fig. 3 that we adapt from H15)}.
{\mtx Their paper includes} multi-frequency coverage from  the
gamma-ray activity  {\mtx lasting} several months to the X-ray and
the optical (V and R) outputs.

{\mtxx In the context of}  our {\mtx previous} discussion, three
points are particularly relevant:

\begin{itemize}

\item 3C 279 is {\mtx one of the 
FSRQs}  detected by Fermi featuring plateau emissions of
different intensity levels lasting months/years;

\item during the period 2013-2014, this source {\mtb produced}
very intense gamma-ray flares
({marked by arrows in our Fig. 3, adapted from H15}) 
{\mtb for which} $L_S / L_{\gamma} \lesssim 10^{-1}$, with no
correlated optical or X-ray enhancements;

\item flare {\mtb "1"} {\mtb of H15 and Fig.~3} at MJD 56646 shows
a {\mtyy particularly} short risetime ($\sim$1 hr) and a
remarkably hard gamma-ray spectrum (photon index $\sim$ 1.6) {\mtb
with no optical counterpart}.
\end{itemize}


\begin{figure}
  \centerline{\includegraphics[trim = 0 0 0 .15cm, clip, width=14.5cm]{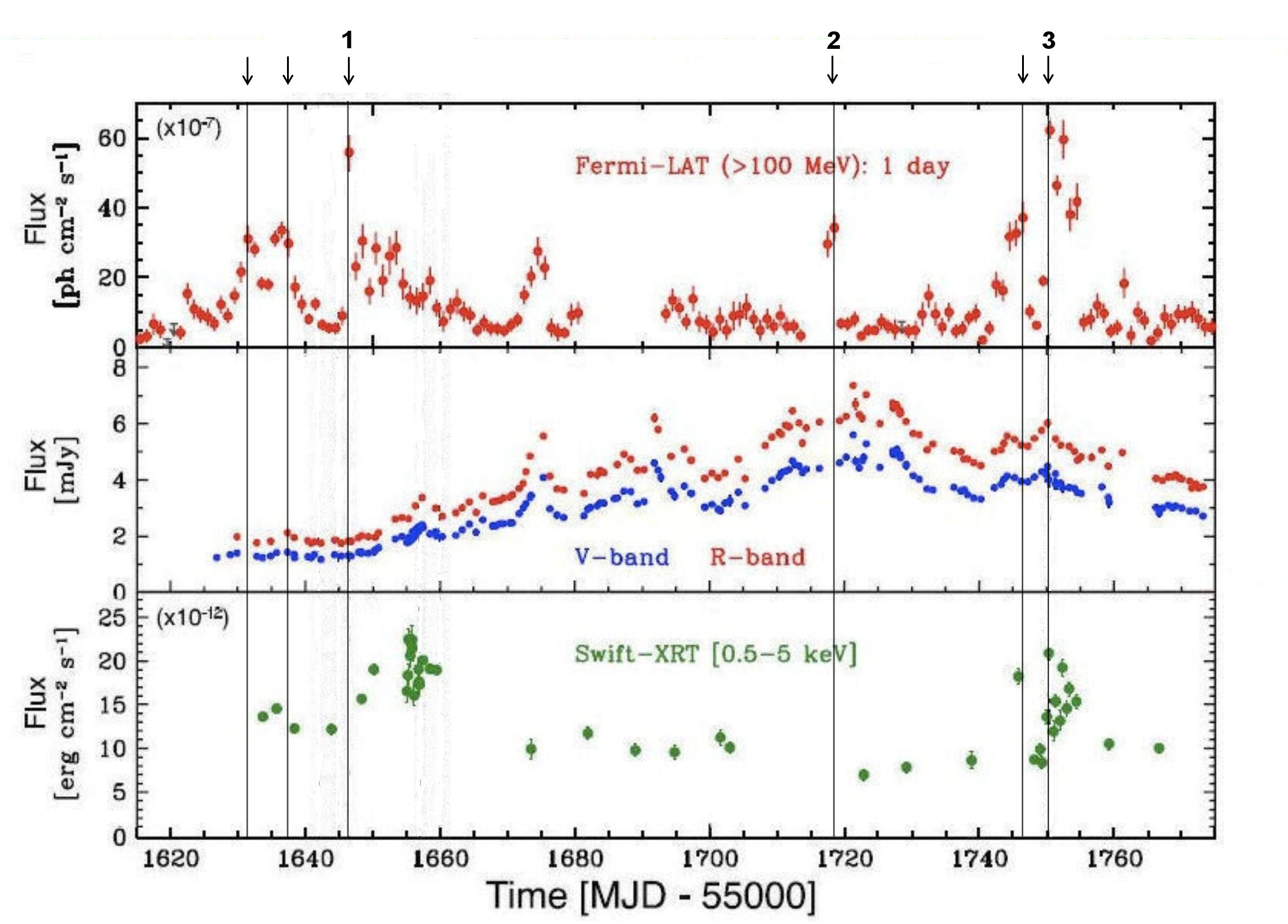}}
    \vspace*{-.5cm}
    \caption{Multifrequency monitoring of 3C 279 (adapted from Hayashida et al. 2015).
    \textit{Top panel:} Fermi-LAT gamma-ray
    lightcurve above 100 MeV. \textit{Middle panel: }optical data in the R-band.
    \textit{Lower panel:} Swift X-ray data. The vertical lines
    highlight the 
    flaring episodes of interest here, i.e., events with
little or no {\mtyy simultaneous} optical/X-ray emissions.
Flares  marked as 1,2 and 3 are labelled in the same way as in H15.}
\label{fig-3c279}
\end{figure}


{\mtb Therefore, we have to acknowledge that {\mtyy a number of}
gamma-ray flares from
{\mtx 3C 279 (as well as {\mtxx those} from other extensively
monitored FSRQs)} do \textit{not} correlate with optical and soft
X-ray events of { comparable} power and time scales. In {many
cases} the former last less and rise more sharply to a much higher
Compton dominance, often on top of a much longer and lower
plateau; they are observed to cover an extended range of {\mtyy
photon} energies from {\mtyy 100 MeV to some 10 GeV}.


The combination of these features is 
{\mtxx \textit{beyond} the  reach, and
  even against the predictions},
of  {\mtx the} canonical one-zone source structure that by
construction  {\mtx yields} highly correlated S and EC radiations.
In addition, to produce  intense EC  this  simple structure
requires a dense and extended  bath of seed photons that easily
{\mtx could} absorb by pair production the high energy radiation
after emission.
Thus  we are led to investigate 
 {\mtx alternative
emission scenarios; we discuss next a "mirror" process,
{\ff possibly } the simplest variant of one-zone modelling.}


\begin{figure}[t!]
  \centerline{\includegraphics[width=18cm]{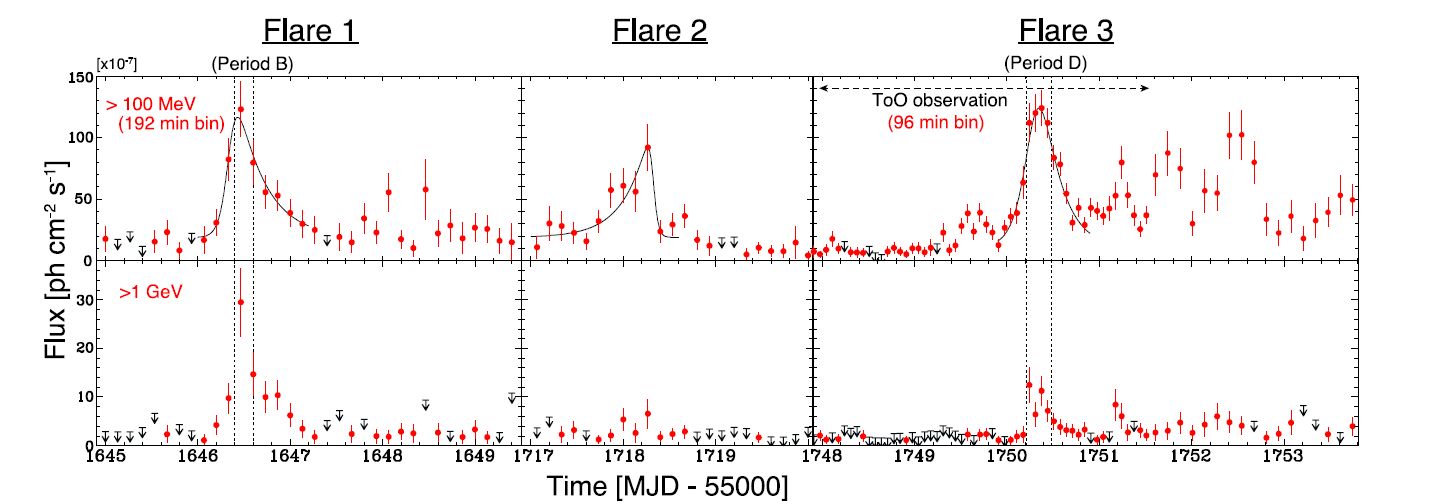}}
  \caption{ Lightcurves observed in flares  1, 2, and 3 of 3C 279
    (see Fig. \ref{fig-3c279}) as reported by Hayashida et al. 2015.}
\label{fig-flares}   \vspace*{1.cm}
\end{figure}

\section{A viable alternative: mirrors}

{\mtx In our approach,} a "mirror" is provided  by  
{\mtxx an individual} plasma cloud or clump with number density $n
\sim 10^6$ cm$^{-3}$ and size $r_m \sim 10^{16}$ cm {\mtxx located
within
the jet 
opening angle}.  {\ff Such a cloud} reprocesses and scatters back
{part of} the impinging S radiation, {emulating a} "reflectivity"
$f \sim 0.1$. Such a value is {\mtx
large} enough to provide a relevant 
{\mtxx enhancement} of seed photons
{\ff before }  the  mirror. Mirrors located within the BLR {\mtxx
or near its boundary} have been discussed by a number of authors
(Madau \& Ghisellini 1996, B\"ottcher \& Dermer 1998, B\"ottcher
2005, {\ff V14, Aliu et al. 2014}); a location beyond the BLR
{\mtx was} recently advocated by V14 who focused on the
observations of 3C 454.3.










{\mtx The} action of a mirror \emph{outside} the BLR  is indicated
by a number of {\mtx conditions required to explain the
observations}:  strong and fast EC flares {from a large but}  {
localized} density of soft photons;
production  of gamma-ray flares uncorrelated with comparable
emissions in the optical/X-ray bands; flares standing on top of a
longer plateau;  unabsorbed  {\mtxx high-energy } spectra.
These requirements motivated us to take up the mirror geometry
and  substantially modify it to operate at  distances
\textit{beyond} the BLR.
{\mtxx At these distances, } little gamma-ray absorption is
expected from pair production by interactions with surrounding
photons  so that hard spectra can outgo unscathed.

The geometry adopted {\mtxy by} V14  envisaged a \textit{static}
mirror similar in size and density to a cloud in the BLR,   but
located
at a larger distance $R \sim 10^{18}$ cm.   
{\mtxx Our present } picture contemplates {\mtxx also moving
mirrors, i.e.,
{\mtyy the reflection and emission} resulting from} 
clumps ("plasmoids") that outflow along the jet {\mtxx with
different speeds}. The {\mtxx primary {\ff emission} from a
plasmoid
{\mtyy within} the BLR} can be reflected { back} by a static 
cloud or by preceding {\mtx plasmoids acting as  mirrors} {\mtxx
moving on the same track beyond the BLR region}.
%
%
{ The key feature of our  model is constituted by a confining
\emph{gap} for  reflected seed photons  originated as S {\ff
emission} from plasmoids  carrying their  {share} or retinue of
relativistic electrons. We will see that the  gap  {constitutes} a
transient structure marked  by a narrow width and {\mtbb by }
{short} time scales. Such features break any  detailed correlation
between the slow variations of the large-scale magnetic field
governing the S emission, and the build up  of the  localized
photon density inducing the EC process.}
We discuss below  how such a 
{\mtx scenario}  can meet all of the above requirements.


\subsection{A static mirror}


{\mtx In the static mirror case, a  partially reflecting cloud is
assumed to  
{\mtxx cross the jet's}  radiative cone at a distance $R_m$ from
the central BH.}
 The mirror reflects back a
fraction $f \sim 10\%$ of the S photons emitted  by active
plasmoids as they travel with the jet. {\mtxx If the plasmoids
share {\mtyy with the jet} a} bulk Lorentz factor $\Gamma \sim
10$, they will approach the mirror at a relativistic speed $\beta
\, c = c \, (1- \Gamma^{-2})^{1/2} \simeq 0.995 \, c$.

The photons  start out from the emitter when this is well below
the position  of the mirror, at a distance {$R_1 \ll R_m$ from the
central BH,  where the magnetic field is still large enough for
intense S emission to occur}  {(see the lower part of Fig.
\ref{fig-basics})}. After reflection, {\ff photons} travel back to
meet again the advancing emitter, and are confined {\ff to} the
narrowing gap between it and the mirror ({see the upper part of
Fig. \ref{fig-basics}}). {In simple form,} the relativistic
travel-time condition for this to occur is given by $R_m - R_1 - d
= \beta (R_m - R_1 + d)$, see 
B\"ottcher \& Dermer (1998) and V14. This  {leads to evaluate the
gap size} $d$  {\ff in the limiting form }
 \be  d \sim \frac{R_m}{4 \, \Gamma^2} \label{eq-1} \en
\noindent in the  
{\mtxx laboratory} frame.  As the plasmoid approaches the mirror,
the gap shrinks down to {\mtxx match} the mirror size;
{\mtxy so, it } 
{\ff attains} a few times
$10^{16}$ cm, with a related time scale $d/c$ of a few days.

 {Consider now that} the energy density in the gap, as
experienced by the advancing emitter  plasmoid (with the
associated relativistic electrons),  is {\mtdd
\textit{mirror-boosted} both} on the outward and on the inward leg
of the photons' journey. So it {\mtbb reads} $ U'_m =  \eta \, U_m
\, \Gamma^2 $ (see Eqs. \ref{eq-98} and \ref{eq-99} in Appendix B
and Appendix C), that is,  \be U'_m = \frac{f \, \eta}{2 \, \pi \,
c} \, \left( \frac{r_m}{R_m} \right)^2 \, \frac{L'_S \,
\Gamma^4}{d^2} \, \Gamma^2 \, . \label{eq-90} \en \noindent {\mtbb
Here $\eta$ is an angular factor of order unity ({\mtxx detailed}
by Eq. \ref{eq-gtilde} in Appendix C)} and $L'_{S} \sim 10^{42 }
\, \rm erg \, s^{-1}$ is the co-moving S luminosity from a
plasmoid near $R_1$,  corresponding {\ff in the usual isotropic
evaluation} to an observed  value $L_S = L'_{S} \, \Gamma^4 \sim
10^{46}$ erg/s. {\mtdd Note that the process of mirror-boosting
introduces {\mtxx the} additional factor  $\eta \, \Gamma^2$ in
the expression for the comoving energy density $U'_m$. Indeed, Eq.
\ref{eq-90} shows that $U'_m$ has a dependence on $\Gamma$ that
results from the energy density as { seen by the mirror} (the
factor $\Gamma^4$), times a
 {further}  factor $\eta \Gamma^2$ induced by mirror
reflection.}

The photon {energy} density in the gap as given by Eq. (2) {\mtbb
induces} a flare of up-scattered  EC {\mtbb gamma-ray emission}
that grows as $1/d(t)^2$ {\ff to yield} a sharp spike.
 {The {\ff specific time decrease } of $d(t)$ is given by Eq. \ref{eq-distance}
  in
Appendix A, and the resulting rise of the flare is represented in
Fig. 5.}. {\ff  Eventually, $d(t)$ narrows  down to match the
mirror size $\ff {r_{m}}$ and the gamma-ray intensity {\ff
attains} its peak. Using the travel-time condition (Eq.
\ref{eq-1}) to
 {expose the overall dependence on $\Gamma$}, we obtain the peak value} \be
U'_{m,p} = \frac{16 \, f \, \eta}{2 \, \pi \, c} \, \left(
\frac{r_m}{R_m} \right)^2 \, \frac{L_S}{R_m^2} \, \Gamma^6 .
\label{eq-91} \en
%
%
%
{\ff Note that similar high powers of $\Gamma$ also arise in {\ww
similarly anisotropic} conditions implying head-on photon-electron
collisions as discussed e.g., by Dermer \& Schlickeiser 2002,
Sikora et al. 2009, Ghisellini \& Tavecchio
2009}.  

{\ff From Eq. 1, {\ff the timescale} for the lightcurve rise} in
the observer's frame turns out to be
\be T \simeq \frac{R_m}{4 \, c \, \Gamma^4}, \en that is, a few
hours. {\ff After having reached its peak, the lightcurve may fall
abruptly if the seed photon 
{\ff replenishment} is cut off. Such a condition occurs when
plasmoid and mirror merge (as the approaching plasmoid screens out
the mirror from the incoming photons), {\ff and/or when the
plasmoid sidesteps the mirror and relativistic de-boosting
applies}. {\ff Such conditions {\ww occur on an observer's
timescale of
order $\sim  r_m / 2 \, c \, \Gamma^2 \sim \rm 1.5 \, hr$, and may} yield} 
strongly asymmetric flares as {\ww observed in 3C 279 (flare 2 of
{\ff Figs. 4 and 5})}.}

{\mtxy {\mtyy As to} the gamma-ray luminosity $L'_{\gamma}$, one
has to account also for the emitting volume with its possible
dependence on $\Gamma$. In fact, if the volume comprises just the
plasmoid and its close vicinity, it will not depend on $\Gamma$
and the scaling of $L'_{\gamma}$ will be $L'_{\gamma} \propto
\Gamma^6/R_m^2$, {\mtyy as  given by {\ff the limiting }  Eq. 3}.
%
On the other hand, if the emitting volume involves the whole gap
and reads $\pi \, r_m^2 \, d'$ (where $d' = d / \Gamma$ is the
comoving gap width), it will start from large values and
eventually shrink to $ \pi \, r_m^3 / \Gamma $, so that
$L'_{\gamma} \propto \Gamma^5$ holds. In either case,
%
the interesting possibility arises of bright and short gamma-ray
flares induced \textit{outside} the BLR.}

{ It is also important to {\mtcc check}
{\mtyy in the plasmoid  frame} the source opacity, $\tau'_{\gamma
\gamma} \sim n'_{ph} \, (\sigma_T/3) \, l'$  due to pair
production by photon-photon interactions; {\mtcc here $\sigma_T$
is the Thomson cross section, and $l'$ a typical {\mtcc
integration} length. Pair production against the {\mtcc observed }
GeV photons  occurs for
 {comoving } photon energies (1 GeV)/$\Gamma$ against softer
photons of energies  $\epsilon' \simeq 10$ keV, that are
{\mtyy reflected} by the mirror at energies $\epsilon \simeq 1$
keV. Denoting by $\alpha_{\rm keV} \lesssim 10^{-1}$ the fraction
of mirror-reflected luminosity  {in the latter range,} we have
\be \tau'_{\gamma \gamma}= \frac{f\, \eta \, \sigma_{T} \,
\alpha_{\rm keV}} {6 \, \pi \, c \, \epsilon' } \, \left(
\frac{r_m}{R_m} \right)^2  \, l'  \, \frac{L_S}{d^2} \, \Gamma^2 =
\frac{f\, \eta \, \sigma_{T} \, \alpha_{\rm keV}} {{ 6 \, \pi} \,
c \, \epsilon' } \, \left( \frac{r_m}{R_m} \right)^2  \, l'  \,
\frac{L_S}{d'^2}   \en where in the last expression we have used
{\mtxy again} $d' = d/\Gamma$. {Over a comoving } integration path
$l' = d'$, we have \be \tau'_{\gamma \gamma} = \frac{f\, \eta \,
\sigma_{T} \, \alpha_{\rm  keV}} {6 \, \pi \, c \, \epsilon' } \,
\left( \frac{r_m}{R_m} \right)^2 \, \frac{L_S}{d'} \simeq 2 \cdot
10^{-4} \, \frac{L_{S,46}}{R_{m,18}} \, \frac{\alpha_{\rm
keV}}{0.1} \;
\Gamma^3 ,\en where we 
{\mtx adopted} $r_m/R_m = 1/30$, $L_{S,46} = L_S / (10^{46} \, \rm
erg \, s^{-1})$, and $R_{m,18} = R_m / (10^{18} \, \rm cm)$.
For $\Gamma \sim 10$, {\mtyy in the gap} {\mtcc we find } $
\tau'_{\gamma \gamma} \lesssim 1$. Whence {\ff we obtain the upper
limit for the static mirror distance
$  R_m \simeq 5 \cdot 10^{17} \, \rm cm $ within which pair
absorption is significant.} For {\mtdd larger distances, the
optical depth for pair production } is less than unity.









{\mtyy Such a mirror geometry} can explain in simple terms
 the puzzling behavior of 3C 454.3 in November 2010, and also that of
3C 279 at the peak of the lightcurve  marked as "flare 2" by H15
(see our Fig. 3), including the absence of comparable correlated
enhancements of the S emission in the optical band  {and the hard
spectra observed to escape from the source.}

A {\mtv static} mirror of reflectivity $f \sim 10 \%$   may be
provided by a stray  BLR cloud hit by the jet (V14), or  by the
atmosphere of a red giant star crossing the jet (Khangulyan et
al., 2013), or more likely by a {\ff lagging} plasmoid left over
in the jet by a previous {\mtyy  ejection} episode. We favor the
latter possibility since it makes  {easier}  for a plasmoid to
pick up a suitable mirror,  and also because the {observed}
flickering plateau apparently requires - in addition to the large
emitter plasmoid originating the strong spike - a whole {\mtv
string} (a "train" in the terminology of V14) comprising many
smaller companions.


\subsection{A moving mirror}

  {The {\mtxx moving-mirror} scenario  envisages}  successive
plasmoids formed and ejected along the jet at different times with
different {\mtxx speeds}. 
A plasmoid moving outwards at a velocity  $c \beta$ as discussed
in the previous subsection may find  on its track another, {\mtv
lagging member of a string} previously ejected,  cruising at a
slower speed $c \beta_o < c \beta$. Then the back side of the
preceding plasmoid can provide a moving mirror for the following
one. {\mtxx This} condition is analyzed in detail in the Appendix
A, and the main results  are reported here {(see Fig.
\ref{fig-basics} for details and definitions)}. {\ff For
simplicity,
we focus here on the 
representative case of constant speeds, and postpone the case of
time variable speeds to a forthcoming publication.}
%

Recall first that {\mtxx in  a static mirror geometry} {\ff the}
photons emitted at {\ff position $z_1$ and time $t_1$} propagate
outwards over a distance $x_1$ to {\ff a}  mirror position
$\tilde{R}_m$; then they are reflected back to meet the emitter
after propagating over an additional {\ff distance $x_2$}. Meeting
at the point $z_2$ is governed by the travel-time condition that
rewrites as $ z_2 - z_1 = \beta \, (x_1 + x_2)$
after B\"ottcher \& Dermer (1998),  where we used the {\mtx
shorthands}: $z_1 = R_1 $, $z_2 = \tilde{R}_m - \tilde{d} $, and
the definitions $x_1 = \tilde{R}_m - R_1  $, $ x_2 = \tilde{d}$
(see Appendix A and Fig.~\ref{fig-basics}). In the case of a
moving mirror, we have to consider also the speed $\beta_o$ of the
mirror itself, in addition to the speed $\beta$ of the plasmoid
emitting the {\mtv primary} S radiation.



Here it is convenient to  define the confining  \textit{gap} as
the distance $\tilde{d}$ between the mirror location $Z(t_r)$ at
the reflection and the  {position}  $z_2(t_C)$ reached  by the
plasmoid at {\ff the time $t_C$ of } the Compton {\ff radiation}
{\ff (which provides  the abscissa 
 $t$ of the representative
gamma-ray lightcurve plotted in Fig. 5);}
%
{\ff this yields } $ \tilde{d} = Z(t_r) - z(t_C)$ which is
calculated (see Eq. \ref{eq-tildee})   to read \be \tilde{d} =
D_{in} \left[ \frac{(1-\beta)}{(1 + \beta)(1 - \beta_o)}
\label{eq-tilde4} \right] . \en
%
{Here the} quantity $ D_{in} = \Delta z + \beta_o \,  c \, (t_1 -
t_o) \equiv \Delta z + D_o $ is the sum of two components: (1)
{\ff an initial distance  $\Delta z = \tilde{R}_m(t_1) - R_1(t_1)$
between the  moving mirror at radius $\tilde{R}_m$ and the point
$R_1$ where the emitter radiates}; (2) a kinematic mirror-plasmoid
initial distance $D_o = \beta_o \, c \, (t_1 - t_o) $ travelled by
the mirror
{\ff during } the time interval $t_1 - t_o$.

Eq. \ref{eq-tilde4}  constitutes our main \textit{kinematic}
result for the case of a moving mirror.  For a static mirror with
$\beta_o = 0$ it goes into Eq. 1; on the other hand, for $0 <
\beta_o < \beta$ the distance $\tilde{d}$ {may be approximated as}
\be \tilde{d} \simeq \frac{D_{in}}{2} \,
\left(\frac{\Gamma_o}{\Gamma} \right)^2 \label{eq-tilde3} .\en
%
Thus   the distance between the  mirror advancing at a speed
$\beta_o \, c$ and the plasmoid catching up with a speed $\beta \,
c$ can {\mtxx be made } substantially {\textit shorter} than
$D_{in}$, at the time of the inverse Compton up-scattering of the
reflected photons. {\mtv In terms of the \textit{relative} Lorentz
factor $\Gamma_r \simeq \Gamma/2 \, \Gamma_o$ (the simple
expression that holds for $\Gamma_o^2 \gg 1$, $\Gamma^2 \gg 1$,
and $\Gamma > \Gamma_o$, see Appendix C) Eq. \ref{eq-tilde3}
writes in the form $\tilde{d} \simeq (D_o / 2) \, /(4\,
\Gamma_r^2) $ } analogous to Eq. 1.

The  energy {density} ({in the} plasmoid comoving frame) of the
reflected photons  that trigger Compton {scattering } writes as
(see Eq. \ref{eq-100}, {and Appendix C }for details)
\be U'_m =  \tilde{\eta} \, U_m \, \Gamma_r^2 =
 \frac{f \, \tilde{\eta} \, L'_S}{ 2 \, \pi \, c} \, \left( \frac{r_m}{D_o}
\right)^2 \, \frac{1}{D_o^2} \left( \frac{\Gamma}{\Gamma_o}
\right)^8  \, \Gamma_r^2 , \label{eq-ummm} \en
{\mtcc where $\Gamma_r$ is the  Lorentz factor of the plasmoid
\emph{relative} to the moving mirror,}  and Eq. \ref{eq-tilde3} has
been used for the gap size {\mtv for $\Delta z = 0$}.}
  In Eq. \ref{eq-ummm}
 we denoted  with  $L'_S$ the S luminosity emitted at the height $z_1$ by  the plasmoid in its
comoving frame.
%
{\mtcc Values that  reasonably apply to a moving mirror-plasmoid
configuration {\mtv of interest here} are $r_m \simeq 3 \cdot
10^{16} \, $cm, $D_o \simeq 3 \, r_m$, $f = 0.1 \, f_{-1}$, and
$\Gamma_o = 3$}.  {Thus}  we have {\ff the limiting value} \be
U'_m \simeq (f_{-1} \, 10^{-7} {\rm \; erg \, cm^{-2}}) \, L'_{42}
\, \frac{1}{ (\Gamma_o/3)^2} \, \left( \frac{\Gamma}{\Gamma_o}
\right)^8 \, \Gamma^2 \label{eq-u'2}, \en where $L'_{42} = L'_S
/(10^{42} \rm \, erg \, s^{-1})$.

{\mtxx For a given number of energetic  electrons,}
 it is 
{\ff {\ff useful} to {\ff refer to a} }
   typical
photon energy density {\mtv around a }  plasmoid moving {\mtdd
through} the {\mtv  photon bath of the } BLR, that {\mtv reads}
\be U'_{BLR} =
\frac{4}{3} \, \frac{ \xi_{BLR} \, L_D \, \Gamma^2}{4 \, \pi \, c
\, R_{BLR}^2 } \simeq (0.03 \; {\rm erg \, cm^{-3}}) \, L_{D,46}
\, \frac{\xi_{BLR}}{0.1} \, \Gamma^2 . \label{eq-ublr} \en
\noindent {Here we used the disk luminosity} $L_{D,46} = L_D
/(10^{46} \rm \, erg \, s^{-1})$, {\mtcc the radius } $R_{BLR} = 3
\cdot 10^{17} \rm \, cm$, {\mtdd and the average BLR covering
factor $\xi_{BLR} = 0.1 $.}
{\ff The value of $U'_{BLR}$ rapidly decreases outside the BLR,
and there would be {\ff little} seed photon density available for
inverse Compton scattering, were it not for the mirror mechanism.}
{\ff Outside the BLR,}  
 $U'_m$ can exceed the 
value $U'_{BLR}$ {\ff previously encountered in the BLR } for a
{\mtcc range} of plasmoid kinematic and radiative conditions;
{\mtcc so the effectiveness of a mirror}
in producing gamma-ray flares can vary. An interesting situation
{\mtcc arises}  when $U'_m$ is comparable to, or larger than the
{\mtcc BLR} value. {\mtdd In fact, }
 $U'_m \geq U'_{BLR}$ {\ff obtains}  for \be \frac{\Gamma}{\Gamma_o} \gtrsim 4 .
\en
{Accordingly}, an {\mtv interesting} configuration for the
moving
mirror-plasmoid geometry {\mtcc  is obtained with} 
the following parameters: $ \Gamma \simeq 10 - 15, \Gamma_o \simeq
2-3, r_m \simeq 3 \cdot 10^{16} \, \rm cm$, {\mtdd and a location
beyond the BLR radius}.

Thus a moving mirror geometry can provide  conditions
\textit{internal} to the {\mtv jet} {\mtb for}  a substantial
enhancement of the local photon bath at a considerable distance
from the BLR. In fact,  in this geometry the distance where the EC
radiation can occur and constitute  a strong gamma-ray flare with
(isotropized) $L_{\gamma} \sim 10^{46} \, \rm erg \, s^{-1}$ is
\be \delta z = z_2 - z_1 = 2 \, \beta \, D_{in} / (1 + \beta) ( 1
- \beta_o) \simeq 2 \, D_{in} \, \Gamma_o^2 \, , \en which can  
substantially {\mtcc exceed}   $R_{BLR}$.
{\mtv For}  reference values $D_o \simeq 3 \, r_m $ , $ r_m \simeq
3 \cdot 10^{16} \, \rm cm$ and with  $\Gamma_o = 2$, we have \be
\delta z \simeq 7 \cdot 10^{17} \, \rm cm , \en {\mtbb while for
$\Gamma_o = 4$ we have} \be \delta z \simeq  3 \cdot 10^{18} \,
\rm cm . \en
{An {effective} mirror mechanism for gamma-ray production can
therefore operate just outside the BLR boundary or somewhat
beyond, depending on jet conditions. On the other hand, an upper
limit $ z_1 < 0.3 $ pc to the distance within which the plasmoid
efficiently {\mtv radiates by EC} is set by the condition that
below the mirror the magnetic field be high enough to {produce} a
primary, comoving S radiation $L'_S \gtrsim 10^{42}$ erg/s  to
illuminate  the mirror.}

To summarize:  plasmoids of sizes $r_m \sim 3 \cdot 10^{16} \, \rm
cm$ with somewhat wider spacings {\mtxx (condensed by a jet
instability to form a string of a few dozens)} can emit, partially
reflect and transiently confine 
{\mtyy S}~photons to produce intense and short Compton flashes.
The primary power $L'_S$ can be {\mtv fed} as the emitter sweeps
the surrounding {\mtbb average} magnetic energy at a rate $\pi\,
r^2_m\, \, \beta\, c\, B^2/8\pi \sim 10^{42}\, B^2$ erg/s {\ff
(with $B$ in Gauss)}, {provided that such a } {\mtx power} goes
into replenishing the electron energies  up to { values}
$\gamma \gtrsim 10^3 $.
A fitting  physical process to embed both a prolonged electron
acceleration and  the fragmentation/coalescence {\ff process} of
jet plasma into plasmoid strings is suggested on noting  that fast
gamma-ray flares often arise {\mtv near} the maxima of mild, slow
enhancements of S emission in the optical - soft X-ray bands (see
Fig. 2 for the case of 3C 454.3). In other words, { mild}
large-scale growth of the magnetic field apparently paves the way
to  {bright} small-scale activity at high energies.

Such conditions match those expected from  a scenario of magnetic
reconnections in a collisionless plasma (see Kagan et al. 2015,
who review classic theoretical work and their own recent numerical
simulations); {\mtv the process} is started where large-scale
field reconfigurations in the jet set layers of opposite,
annihilating {\bf B} lines. The magnetic energy so liberated feeds
kinetic tearing instabilities that lead  to {\mtxx local} jet
\emph{fragmentation} into strings of {\mtcc separate } plasmoids.
{\mtcc These include} {a few giant ones} arising from repeated
cycles of coalescence and condensation of smaller companions. The
cycles are to take the fragments from the minimal, {\mtcc kinetic}
scales
set by the inertial skin depth {\mtcc at} $10^2\, c/\omega_p $ 
{in terms of the electron plasma frequency} $\omega_p \simeq (5
\cdot 10^4 \; {\rm s^{-1}}) \, (n/\gamma)^{1/2}$ {\ww (where $n$
is the electron number density in units of $\rm cm^{-3}$)}, {up to
}the sheath overall length $2\, \ell \sim $ several light-days
{\mtb (see Fig. \ref{fig-basics}}).


Meanwhile, the ensuing electric fields efficiently
\emph{accelerate} electrons up to { energies} $\gamma \gtrapprox
10^3$ in, or between the plasmoids even under initial conditions
of moderate magnetization $\sigma = B^2/\Gamma \,n\, m_p\, c^2
\sim 10 $. Losses by S cooling and  by adiabatic expansion are
replenished by
 {continuous acceleration going on} as long as the process
of repeated fragmentation and coalescence  {lasts} in the magnetic
reconnection sheaths. {\ff Alternatively, magnetic reconnection
has been considered in the context of forming  mini-jets (e.g.,
Kagan et al. 2015). However, Narayan \& Piran 2012 discuss the
 considerable relative Lorentz
factors of the mini-jets and their large multiplicity, which
concur to require a quite large power from the central source.}

\section{Specific marks of a new class among the gamma-ray flares of 3C 279}

We have discussed   why {\mtb in the  {framework} of the canonical
one-zone model} a S - EC source located in the BLR can account
 {neither} for  luminosity ratios $L_{\gamma}/L_S >> 1$
 {nor} for the lack of {\mtv detailed} correlations between the gamma-ray
flares and the synchrotron optical and soft X ray components.  It
is clear today that a number of gamma-ray flaring episodes from
prominent FSRQs show features  that are challenging such {\mtb a
framework}. Although {\mtg a} comprehensive study and
interpretation of
 {all types} of gamma-ray flares is beyond the scope of the
present paper, we will briefly  {place} our alternative mirror
model  {in} the context of the 2013-2014 activity of 3C 279 that
provides a 
collection of diverse {\mtv flare prototypes.}

\begin{figure} 
 \vspace{-4.5cm}
   \centerline{\includegraphics[width=14cm]{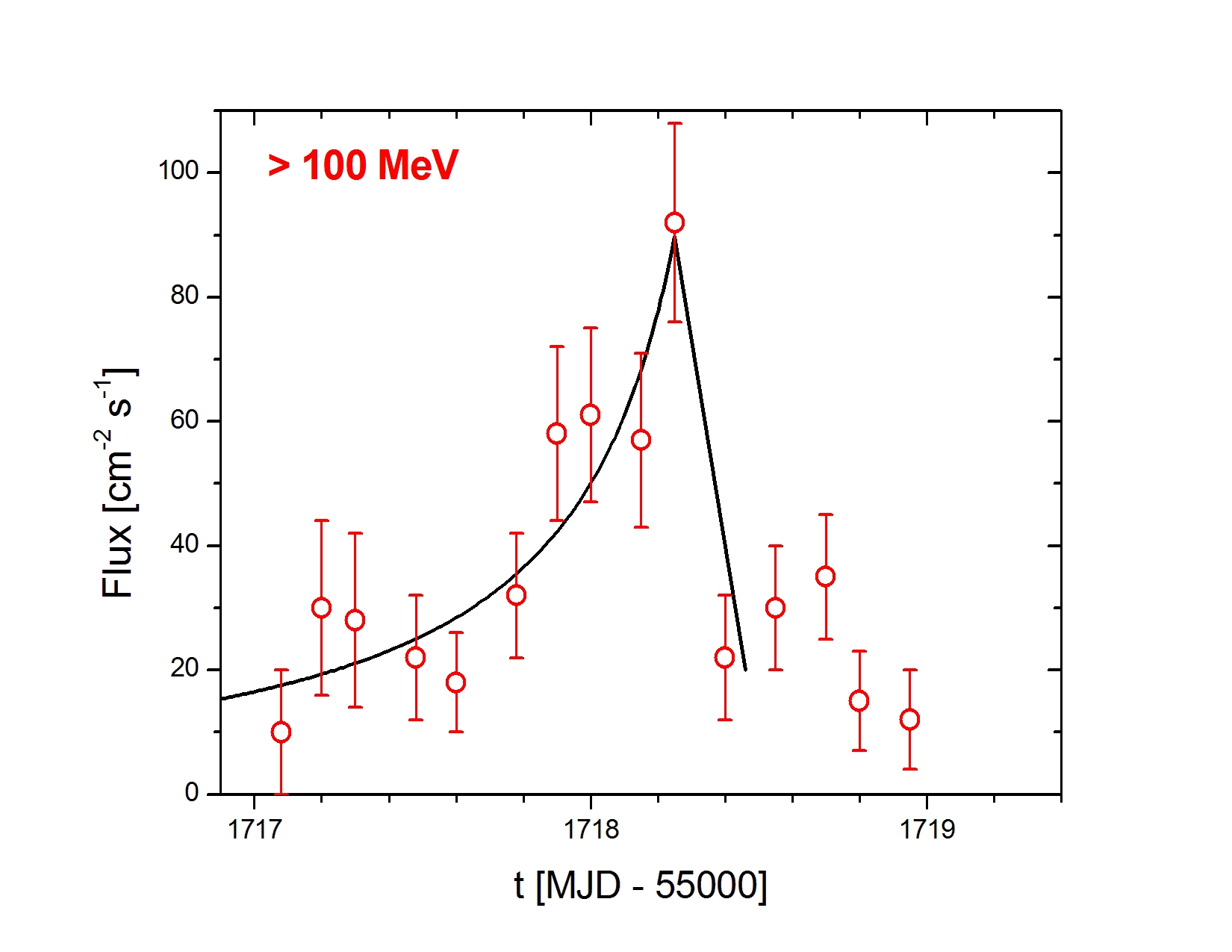}}     
\caption{ {\ff  Gamma-ray lightcurve of flare 2 of 3C 279 (data
points from Hayashida et al. 2015, zoomed from Fig. 4)
superimposed to the
 gamma-ray emission calculated from the mirror-driven mechanism
discussed in the text (solid curve). The rise is produced by the
seed photon density increasing as $U'_m \propto 1/d^2$ (see Eq. 2)
with the gap size $d(t)$ shrinking after the linear relation given
by Eq. \ref{eq-distance}. The {\mtyy sharp} fall follows from the
sudden decrease of the seed photon density when the emitter
crosses or sidesteps the mirror {\ww on a timescale shorter than
the data resolution (193 min., H15); see Sect. 3.1 for
discussion}.}} \label{fig-theory}.
\end{figure}

We based  our considerations on the recent observations {\mtv of
3C 279} by H15; in our Fig. \ref{fig-3c279} we highlight from that
paper a number of gamma-ray episodes relevant to our analysis. The
vertical lines mark several gamma-ray flares with little or no
simultaneous optical and X-ray enhancements. H15 identified three
particular gamma-ray flares of interest;  we 
{\mtg singled out} in our  Fig. \ref{fig-flares} the lightcurves
of these events in the range from  0.5 to 10 GeV.

We first {focused} on "flare 2" that features  a markedly
\emph{asymmetric}  lightcurve.  Here the rise  over $\sim$1 day is
followed by a peculiarly sharp fall in less than one hour. This
behavior was noted  by H15, who parametrized the rise-time and the
decay time at 6.4 hr and 0.6 hr, respectively. Such a peculiar
shape is not often observed in high-energy sources, and points to
a specifically time-asymmetric emission mechanism. We {identified}
such a behavior with that expected from Compton up-scattering of
radiation reflected by a moving mirror, as we have computed and
shown in our Fig. \ref{fig-theory}.

{From our kinematics we obtained} steeply rising
spikes from EC triggered by a doubly \textit{boosted} photon
density; this is  emitted  {at about the height of the BLR,   and
then
 {}reflected back} {into the gap between the outer mirror
and the advancing} plasmoid (see Eq. \ref{eq-u'2} with its strong
dependence on $\Gamma$). Eventually, {\mtv as the plasmoid pastes
on, or sidesteps the mirror,  the seed photon density decays
sharply as the mirror-boosting no longer applies}.  Fig.
\ref{fig-theory} shows {the sharply asymmetric pattern expected
for  the lightcurve produced
by a mirror process  with the simplest 
kinematic plasmoid-mirror relationship {\mtg as discussed} in
Sect. 3. Such a \emph{kinematically} triggered and quenched
radiation pulse agrees with the observations of "flare 2" in 3C
279 reported by H15}. {\ff  The detailed agreement  between flare
2 and the asymmetric gamma-ray lightcurve produced by our simple
mirror mechanism is shown in Fig. 5.}

{On the other hand, more complex conditions may arise} from
interactions between mirror and plasmoid as the two get {\mtxx
close} to near contact, that is, to a distance smaller than $r_m$.
{For one,} {\mtxx when we also account for mirror motion outwards,
Eq. 10 shows that the spike is softened to a more rounded peak as
the gap
shrinks to the mirror size, and ends into  a 
{\mtxx stretched } decay. In addition,} {\mtcc bright  EC
radiation implies severe electron cooling,} while forced electron
re-acceleration may occur by compression into the gap of magnetic
field lines dragged along by the approaching plasmoid} in a
delicate balance that is worth discussion. {\mtg The } cooling
time of the energetic electrons that emit gamma-ray photons up to
GeV energies is to be contrasted with the variability time scale.
The inverse Compton cooling time in the comoving  frame is given
by $\tau'_c \simeq 3 \, m_e \, c/ 4 \, \sigma_T \, U'_m \, \gamma
\simeq (1.5 \cdot 10^{4} {\rm \, s}) \, \gamma_3^{-1} \,
\Gamma_1^{-2}
$, where $\gamma_3 = \gamma/10^3$, $\Gamma_1 = \Gamma / 10$, {\mtv
and we used as a reference the energy density of the broad line
region (Eq. \ref{eq-ublr} with $\Gamma = 10$)}.
 Thus the cooling time  turns
out to be  shorter than the {\mtg kinematical} time $ \tilde{d}/
\Gamma \, c \simeq 10^5 \rm \, s$ {\ww in the comoving frame}.
Therefore, the {\mtxx softer decrease of {\ww other observed}
gamma-ray lightcurves} has to depend mainly  on {\mtg other
factors such as
the gap collapse.}
%
%
%
%


{\mtxx  In particular, the slow decays of flares 1 and 3 as
reported in our Figs. \ref{fig-3c279} {\mtg and} \ref{fig-flares}
cannot be accounted for in terms of the associated fast radiative
cooling just discussed\footnote{See also Paliya et al. 2015 for a
recent discussion of flare-3.}. Such slow decays point rather to
the stretching of the lightcurve over times of days that we expect
(as pointed out above) from substantial mirror motion. This is
likely assisted by electron re-acceleration  {\mtxy as} indicated
by the particularly hard spectrum of flare 1, } {\ff see also the
recent detection of photon emission above $\sim 50$ GeV from 3C
279 (Paliya, 2015).}

{\mtg In addition to flares 1, 2, and 3, Fig. \ref{fig-3c279}
shows other gamma-ray flaring episodes that do correlate with
other bands, and so are amenable to standard interpretations in
terms of one-zone source geometry at BLR distance.}
%
%
%
{\mtg 3C 279 is also known for its remarkable gamma-ray flare on
February 2009 observed to correlate with large {\mtg swings of}
optical polarization {\mtg that may be} interpreted in terms of
smooth bending of  magnetic field lines at parsec distances (Abdo
et al. 2010b, {\ff see also Zhang et al., 2015}).}
 at equal
observability/flux. {\mtg We have in preparation}  a detailed
discussion of diverse kinds of gamma-ray flares {\mtg
 between these two extremes.  {\mtyy As anticipated in the Introduction, we expect flares
 belonging to the class discussed here to
 occur in a number of FSRQs including
 3C 454.3, PKS 1510-089, PKS 1830-211, OJ 248.}}


{\mtv To summarize: in our approach the \emph{rise }of a {\mtg
number of} gamma-ray flares
 is dominated by {\mtg localized} kinematic effects, i.e., the shrinking photon
 gap.  On the other hand, the \emph{fall} phase {\mtg may} be
 {\mtxx
 modulated}
 by {\mtxx other} processes. Flare 2 can be
interpreted in terms of a purely kinematical event with a decay
governed {\mtg just} by gap collapse. {\mtg Instead}, flare 1 and
flare 3 show a smoother decay phase, with a particulary  hard
spectrum in flare 1, providing evidence of particle
re-acceleration at work.


%

\section{Discussion and Conclusions}

Strong, sharp flares in gamma rays around  $1$ GeV  are observed
in  beamed blazars of the {\mtx FSRQ} flavor. These sources are
marked not only by a jet-like powerful plasma outflow with a
moderate bulk Lorentz factor $\Gamma \sim 10$, but also by SEDs
dominated by inverse Compton radiation even in {\mtx their low
states of emission}. 
We have discussed {\mtx here} how and why bright gamma-ray flares
\textit{often} do not show {\mtxx clear} time correlation with
emissions in other bands, while they  attain high luminosities so
as to increase the Compton dominance by factors up to $10$.

Such  flares  are widely discussed in terms of the external
Compton radiation process (i.e., inverse Compton up-scattering of
soft "seed" photons),  while the emissions from the optical to the
soft X-ray band are well understood as synchrotron radiation in
the large scale, slowly changing magnetic field $ B\sim 1 $ G that
threads  the jet at distance around $0.1$ pc from the central BH.
Both radiation components are produced by similar or even
coinciding populations of ultra-relativistic electrons 
{\ff with break energies} $\gamma \sim 10^3$ that inhabit the
source.
{\mtx However,}  a S - EC source with the canonical one-zone
geometry cannot account for instances with  luminosity ratios
$L_{\gamma}/L_S \gtrsim 10$ and lack of {\mtx specific} 
correlations 
{\mtx between the synchrotron optical} and 
{\ff the inverse Compton gamma-ray} components.

We found instead  that a mirror-like geometry at distances of
{\mtyy about} $0.3$ pc from the central BH  can yield the
\textit{high} but \textit{localized} and \textit{transient} 
densities {\mtyy of {\ww anisotropic} S photons} needed for a
strong, sharp and short EC flash to occur. {\mtx  This scenario
can
explain} 
 the 
 {\mtx large}  observed gamma-ray luminosity,
short time scales, and related lack of correlations with the
larger scale S emissions. Our geometry is based on S radiation
emitted by individual plasmoids at  {lower heights}  $0.1$ pc in a
string  outflowing with the jet. {Such emission} is
reprocessed/scattered back by an outer mirror at some $R \sim 0.3
- 1 $ pc  provided  by a large, slow plasmoid in a previous
string.

{\mtx A} key feature is constituted by  a narrow and shrinking
\textit{gap} between the mirror and the approaching plasmoid,
where the density of photons is enhanced sharply by double
\textit{mirror-boosting} on their  outward and  inward course. So
{\mtx during time intervals } of hours, the {\mtg photon } density
attains levels high enough to trigger  strong {\mtx flashes} of
high-energy IC radiation. {\mtx Eventually,  this emission} is
saturated and {\mty ultimately} quenched} as plasmoid and mirror
come closer than the mirror radius. {\mtg We emphasize that in
this process the Compton dominance is even enhanced relative to
the already large value marking the FSRQs {\mtx in their plateau
states}.}
Our {\mtx gamma-ray} lightcurves  feature a 
steep rise, and
an even sharper fall
when the plasmoid piles up onto, or
sidesteps the mirror,  so as to suddenly 
{\mtg decrease}  the seed photon {\mtxx density}. On the other
hand, the process allows hard spectra in the range 100 MeV - 10
GeV to escape from the source, since in and around the narrow gap
a low optical depth {\mtg for pair creation by photon-photon
interactions applies. }

\begin{figure}[t!]
\vspace{2cm} \centerline{\includegraphics[width=6cm, angle =
0]{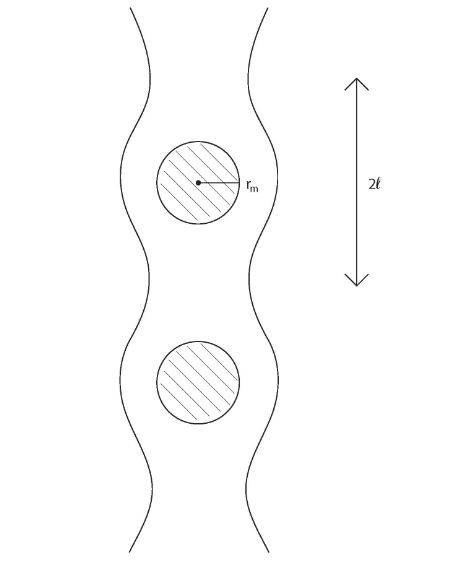}} 
   \caption{\mtx Schematic view along the jet z-axis of the flow
   instability leading to plasmoid formation {\mtyy (see also Kagan et al. 2015)}.}
   \label{fig-plasmoids}
\end{figure}

{\mtyy Our specific source model requires active, emitting
plasmoids with bulk Lorentz factors $\Gamma \gtrsim 10$ to satisfy
the opacity and luminosity requirements; it requires also passive,
mirror plasmoids with $\Gamma_o \sim 3$ to locate the flash of EC
radiation at the outer edge of the BLR. }
{\ff We name  the geometry and emission mechanism discussed in
this paper the "Mirrored-Syncro-Compton" (MSC) scenario for
gammma-ray production in blazars. The MSC mechanism is marked by
strong Compton dominance with no optical/X-ray simultaneous
emission during the gamma-ray flares, by asymmetric flare
lightcurves, and by short timescales.}

Our 
{\ff MSC} source geometry fits in with the 
 prevailing  physical {\mtg view} that contemplates in
the jet a collisionless plasma threaded by large scale magnetic
fields {that include} sheath-like structures; here opposite {\bf B
} lines run sufficiently close over a distance  $2 \, \ell \sim
0.1$ pc to annihilate and convert the magnetic energy into two
forms.
First,  at 
scales of some $10^2$ inertial
skin depths
{\mtxy indicated by the current numerical simulations (see Kagan
et al. 2015)} these plasma/field structures become unstable to
tearing instability, and \emph{fragment} into strings of many
{separate} condensations, i.e., plasmoids, some of which coalesce
into a few giant ones. Thus the macroscopic structure goes through
many cycles of instability-coalescence, with scales up to the
sheath overall length of some $ 0.05$ pc ({see Fig.
\ref{fig-plasmoids}}).
Second, at the particle level the electric fields associated with
annihilating {\bf B} are very efficient in continuously
{accelerating} electrons within and between plasmoids, to attain
high random Lorentz factors { with  energies} $\gamma \gtrsim
10^3$. {\mtxy Ultimately, most of the large radiative powers
observed in such events is driven by the overwhelming bulk energy
flow carried by the jet (e.g., Celotti \& Ghisellini 2008).}

In conclusion,  the radiative and kinematical processes proposed
here can provide the building blocks for \textit{ intense and
fast} gamma-ray flares \emph{uncorrelated} with optical - soft
X-rays events that are most difficult to explain in the context of
the canonical one-zone geometry. This is achieved by locating the
gamma-ray source outside the broad line region; our proposal
accounts for the {\mtxx observed} features marking of a number of
flares: conspicuous increase of Compton dominance;  lack of
{\mtxx time correlation between gamma-rays and optical/X-ray
emissions},
hard spectra {\mtxy produced by re-acceleration and not 
cut off by pair production.}

{\mtg We find the 
{\ff MSC picture discussed throughout 
{\ww the present} paper}  particularly attractive since it offers
three relevant features: it {\ff makes easier for a plasmoid to
{find}}
 a mirror within the jet; large optical/UV photon densities can
be localized just \textit{outside} the broad line region
 so as to  induce \textit{strong} gamma-ray flares in {optically
thin} environments with \textit{no} or \textit{negligible}
correlation with other bands; short timescales can be naturally
produced.} {\mtx The mirror scenario for transient gamma-ray
production presented here can have broad {\mtxx relevance} in
blazar research.}

\vskip .2in

{\bf Acknowledgments: }  We are {\ff indebted to} P. Giommi for
{\ff helpful} discussions and for providing Fig.~2. {\ff We thank
our anonymous referee for useful comments and suggestions}.
Investigation carried out with partial support by the ASI grants
no. I/028/12/0 and I/028/12/2.

{}

\newpage

\appendix

\section{Appendix: Mirror Kinematics}


Let us consider in some detail a geometry in which plasmoid
outflow occurs along the $z$ axis perpendicular to the accretion
disk plane. {\mtxx Fig. \ref{fig-basics} shows the geometry and
the meaning of the relevant quantities.} Let $z_1$ and $z_2$ be
two successive positions of a moving plasmoid with velocity $\beta
\, c$. Photons emitted at {\ff position $z_1$ and time $t_1$}
propagate through a {\ff distance $x_1 $ } to a given mirror
position, and then are reflected back into the same plasmoid after
propagating an additional {\ff distance $x_2 $}; their encounter
with the plasmoid at $z_2$ is governed by the light travel-time
condition \be z_2 - z_1 = \beta  \, (x_1 + x_2) . \label{eq-5} \en
Eq. \ref{eq-5} applies to a static mirror (Boettcher \& Dermer
1998) {\mtxy where we used the definitions given at the beginning
of Sect. 3.2.}
%
We aim at extending the
condition to a moving mirror, i.e., to the case of radiation
reflected back by a plasma clump moving in front of the active
plasmoid. In such a case  two speeds are to be considered:
$\beta_o$ for the reflecting clump, and $\beta$ for the emitting
plasmoid.

We focus on a moving mirror constituted by another plasmoid
able to reflect a relevant  amount of impinging radiation. We also
assume that within the distance span  of  interest
$\beta_o$ and $\beta$ are closely constant. Let us consider
quantities in the observer's frame.
The moving mirror is ejected at the distance $z_1^*$ at time
$t_o$, and moves outward as \be Z (t) = z_1^* + \beta_o \, c \, (t
- t_o) . \en The emitting plasmoid reaches the distance $z_1$ at
time $t_1$ (with $t_1$ larger than $t_o$),  and moves outward
according to \be z(t) = z_1 + \beta \, c \, (t - t_1) . \en {\mt
In the following, we define \be \Delta \,z = z_1^* - z_1 \en In
general $\Delta z \neq 0$, but also $\Delta z= 0$ is viable (see
below for a physical interpretation). } Photons emitted at the
position $z_1$ at time $t_1$ propagate freely according to
$\zeta(t) = z_1 + c \, (t - t_1)$. Reflection occurs at time $t_r$
under the condition $ \zeta(t_r) = Z \, (t_r) $,  that can be
expressed as \be t_r = \frac{t_1 - \beta_o \, t_o + \Delta z /
c}{1 - \beta_o} \label{eq-tr} . \en At the time $t_r$ the moving
mirror reaches the position \be Z (t_r) = z^*_1 + \beta_o \, c \,
\frac{t_1 - t_o + \Delta z/c}{1 - \beta_o} \label{eq-z} . \en
Thereafter, photons are propagating back
and their position along the $z$-axis is given by  $\zeta (t -
t_r) = Z \, (t_r) - c \, (t - t_r)$, for a time $t > t_r$.

The   plasmoid {\mt moving
at height $z(t)$ along the z-axis} is irradiated by the reflected
radiation at time $t_C$, and the travel-time condition reads now
\be \zeta(t_C - t_r) = z(t_C) ,\en that is, \be Z \, (t_r) - c(t_C
- t_r) = z_1 + \beta \, c \, (t_C - t_1) . \en \noindent By using
Eqs. \ref{eq-tr} and \ref{eq-z} we find \be \frac{\Delta z}{c} \,
(1 - \beta_o) + \beta_o \,(t_1 - t_o + \frac{\Delta z}{c}) + t_1 -
\beta_o \, t_o + \frac{\Delta z}{c} + t_1 \, \beta \, (1 -
\beta_o) = t_C \, (1 + \beta) \, (1 - \beta_o) . \label{eq-t2} \en

Eq. \ref{eq-t2} provides the time $t_C$ when radiation hits back
the approaching plasmoid for a given value of $t_1$. As the
emission process is continuous, also the corresponding {\ff
process of mirroring and IC radiation cover a range of $t_C$. We
stress that  the time $t_C$ of  the Compton radiation constitutes
 the abscissa $t$ of the representative gamma-ray lightcurve plotted
as Fig. \ref{fig-theory}.}
%
The highest  energy densities of the reflected radiation obtain at
the time $t_C$ when the shortest  distance is attained between the
moving mirror at $Z(t_C)$ and the  plasmoid following at the
position $z(t_C)$. The mirror-plasmoid distance  at the time $t_C$
of Compton upscattering is
\be d(t_C) = Z \, (t_C) - z(t_C) . \label{eq-t3} \en

Before applying Eqs. \ref{eq-t2} and \ref{eq-t3} in full, we check
them on  the simple case of a static mirror. Then  $\beta_o = 0$
and $z_1^* = R_m$ hold, where $R_m$ is the constant mirror
distance from the central black hole; the reflection time is $t_r
= \Delta z /c + t_1$. The location of the reflected radiation is
$\zeta(t) = R_m - c(t-t_r)$, and the light travel-time condition
becomes $\zeta(t_C - t_r) = z(t_C)$. Thus, we obtain $t_C = 2 \,
(\Delta z / c)/(1 + \beta) + t_1$, and the plasmoid - static
mirror distance $d = R_m - z(t_C) = \Delta z - 2 \, \beta \,
\Delta z/(1 + \beta) = \Delta z \, (1 - \beta)/(1 + \beta)$, where
in this case, $\Delta z = R_m - z_1$.
Thus we recover Eq. \ref{eq-1} 
{\mtv for} $\beta \simeq 1 -  1 / (2 \, \Gamma^2) $.

Having checked our formalism, we  now apply it in full  to the
moving mirror case. The mirror-plasmoid distance as a function of
time is 
given {\ff in the model considered in this paper} by the
\emph{linear} dependence \be d(t) = Z(t) - z(t) = D_{in} - c \, (t
- t_1) (\beta - \beta_o) , \label{eq-distance}\en where we defined
the initial distance at time $t_1$ as \be D_{in} = \Delta z + D_o
,\en in terms of the distance  $D_o$ travelled by the mirror
during the time interval between mirror and plasmoid 
{\mtdd transits through the point } $z_1$, i.e., \be D_o = \beta_o
\, c \, (t_1 - t_o) . \en

From  Eq. \ref{eq-t3}, we derive the mirror-plasmoid distance at
time $t_C$, \be d(t_C) = (\beta_o - \beta) \, c \, \frac{ 2 \,
\Delta z + \beta_o (t_1 - t_o) + (t_1 - \beta_o t_o) + t_1 \,
\beta \, (1 - \beta_o)}{(1 + \beta)(1 - \beta_o)} + \beta c t_1 -
\beta_o c t_o + \Delta z. \label{eq-master} \en \noindent By
adding and subtracting the quantity $\beta_o \, t_1$ and rewriting
Eq. \ref{eq-master} in terms of the initial distance $D_{in}$
we obtain
\be d (t_C) = D_{in} \, \left[  1 + \frac{2 \, (\beta_o -
\beta)}{(1 + \beta) (1 - \beta_o)} \right] . \label{eq-final} \en
Eq. \ref{eq-final} is our main result for the moving mirror case.
We note  that for the case $\beta_o = 0$ (static mirror) we obtain
$d(t_C) = \Delta z \, (1 - \beta)/(1 + \beta) $, again in
agreement with Eq. \ref{eq-1}  {considering that,} {\ff in this
case,} $\Delta z = z_1^* - z_1 = R_m - R_1$.

For both $\beta_o$ and $\beta$ close to unity, we have  \be d
\equiv d(t_C) \simeq D_{in} \, \left[  1 + \frac{\beta_o -
\beta}{1 - \beta_o} \right]  .\en

{\mtb A key quantity } that can be defined as the effective photon
"compression gap" is the distance $\tilde{d}$ between the mirror
location at the moment of reflection $Z(t_r)$ and the distance
travelled by the plasmoid at the Compton up-scattering event,
$z_2(t_C)$, that is \be \tilde{d} = Z(t_r) - z(t_C) ,\en that can
be calculated to be \be \tilde{d} = D_{in} \left[
\frac{(1-\beta)}{(1 + \beta)(1 - \beta_o)} \right] .
\label{eq-tildee} \en For a static mirror with $\beta_o = 0$, we
find $\tilde{d} = D_{in} \, (1-\beta)/(1 + \beta) \simeq D_{in}/ 4
\, \Gamma^2$.

A number of  particular cases may be considered:
\begin{itemize}
\item for $\beta_o > \beta$, the distance $d$ is larger than
$D_{in}$, making the  energy density of reflected radiation
ineffective at inducing an intense inverse Compton radiation;
\item for $\beta_o = \beta$, the distances $d$ and $\tilde{d}$ are
constant and equal 
{\ff a fixed } distance $D_{in} = \Delta \, z = \tilde{R}_m -
R_1$; \item for $\beta_o < \beta$, the {\ff limiting} distances
$d$ and $\tilde{d}$ can be substantially smaller than $D_{in}$,
and {\mtdd read} \be d       \simeq      D_{in} \, \left[ 1 +
\left(\frac{1}{\Gamma^2} - \frac{1}{\Gamma_o^2} \right) \,
\Gamma_o^2 \right] = D_{in} \, \frac{\Gamma_o^2}{\Gamma^2} , \en
\be \tilde{d} \simeq \frac{D_{in}}{2} \,
\left(\frac{\Gamma_o}{\Gamma} \right)^2 \label{eq-tilde2} .\en

\end{itemize}

{\ff We stress that for $\beta_o = \beta$ the mirror mechanism
leads to a constant  flux, while for $\beta_o < \beta$
it yields a time variable high-energy flux.}



\begin{figure}[th!]
\centerline{\includegraphics[width=15cm, angle
=0]{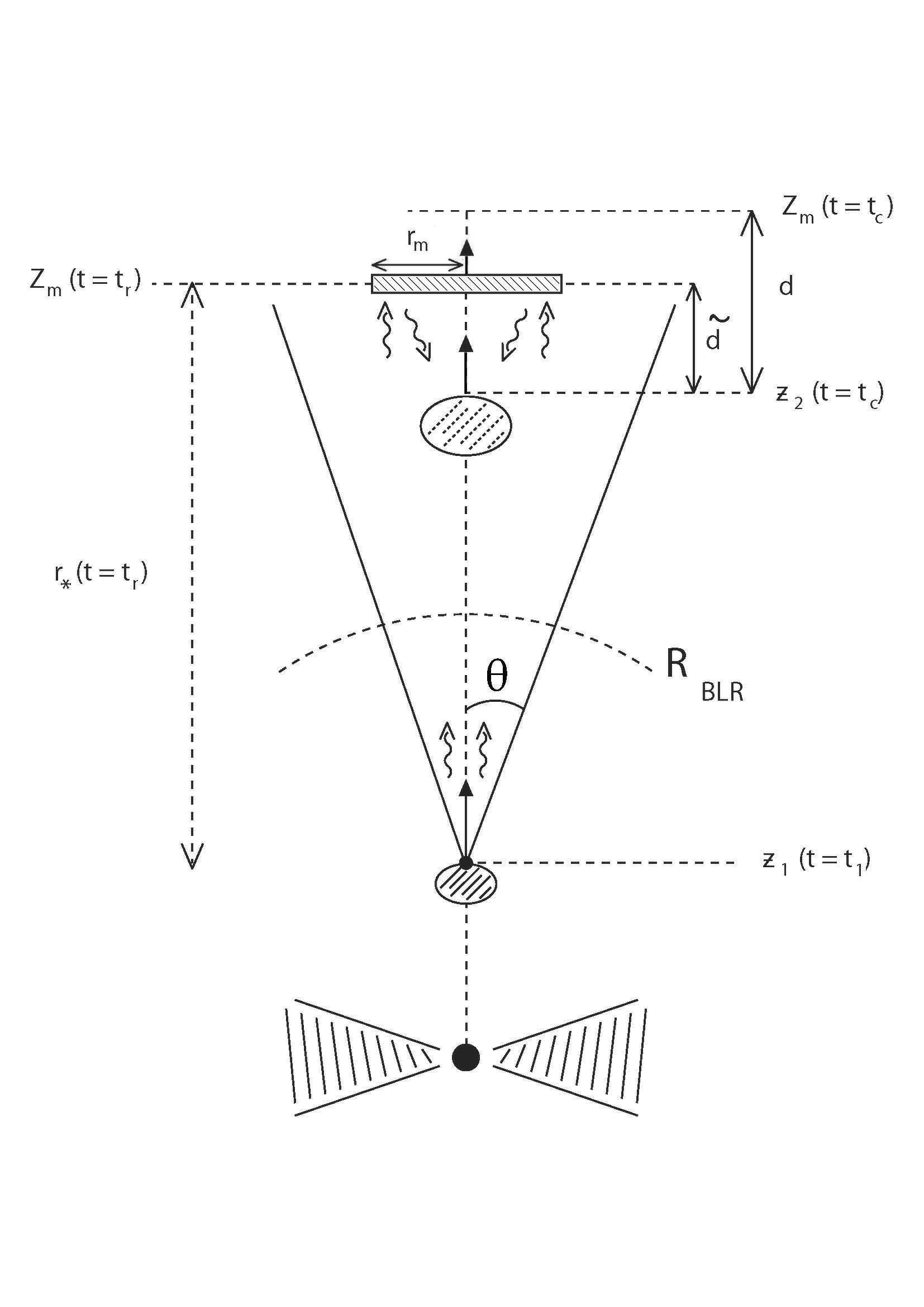}}
\vspace*{-2.cm}
   \caption{Geometry of the kinematic quantities in the  moving mirror case (laboratory frame).}
\label{fig-basics}
\end{figure}


\section{Appendix: Photon Densities}

{ The local photon density in the plasmoid comoving frame $U'$ is
{built up }by different radiative contributions as the plasmoid
moves outward along the jet. Within the BLR, $U'$ is set by the
average soft photon density $U_{BLR} = L_D/4 \, \pi \, c \,
R_{BLR}^2$, so that $U' \simeq  (4/3) U_{BLR} \, \Gamma^2$, where
$\Gamma$ is the plasmoid bulk Lorentz factor, and $R_{BLR} \simeq
3 \cdot 10^{17} \, \rm cm$ ({e.g., Dermer \& Schlickeiser, 1994}).
Outside the BLR, i.e., at distances $r \geq R_{BLR}$ along the
jet, $U'$ is dominated by the contribution of the dusty torus
which has overall luminosity and average photon energies smaller
than those within the BLR. {\ff Compared with the mirrored
synchrotron emission, the latter contribution is less effective in
producing GeV gamma-ray flares by inverse Compton up-scattering of
IR photons also in view of the Klein-Nishina effects}.

A substantial contribution to $U'$ is instead provided by
radiation emitted by the plasmoid within the BLR and only
partially  reflected by a facing mirror at a distance $R_m $
substantially larger than $R_{BLR}$. Notice that the original
radiation flux is only partially captured by the distant mirror at
$R_m$.

Let us first consider a static mirror. Then the fraction of the
plasmoid radiation impinging on the mirror is set not only  by the
flux dilution at the considerable distance between emitter and
mirror, but {\ff is also proportional to} 
the {\ww solid angle $\delta \omega_m \propto \pi \, r_m^2$}
subtended by the mirror of size $r_m$. Thus
 the lab-frame luminosity captured by the mirror is  \be L_m
\simeq L'_S \frac{\pi \, r_m^2}{\pi \, \theta^2 \, R_m^2} \,
\Gamma^2 \,  \label{eq-Lm}\en where the  factor $\Gamma^2$
accounts for the Lorentz transformation of energy and time, and
the emitter is radiates in the solid angle $\pi \, \theta^2$ in
the lab-frame. {\ff In Eq. \ref{eq-Lm} we used the approximation
$R_m - R_1 \simeq R_m$.}
 With  the transformation  $\theta \sim 1/\Gamma$
applied  to relativistic motion, we recover that $L_m \sim
(r_m/R_m)^2 \, L_{obs}$, where $L_{obs} = L'_S \, \Gamma^4$ is the
plasmoid original luminosity in the observer frame. {\ff Eq.
\ref{eq-Lm} assumes $r_m/R_m \leq \theta $ to hold; for $r_m/R_m >
\theta $, the usual relation $L_m = L'_S \, \Gamma^4$ applies. }

The mirror reflects the impinging radiation with efficiency $f$;
at a distance $d$ from the mirror, in the  {observer}  frame the
energy density of radiation reflected back  reads \be U_m =
\frac{f \, L_m}{2 \, \pi \, c \, d^2} .\en {\mtbb In the plasmoid
comoving frame the energy density can be written as \be U'_m = U_m
\, \Gamma^2 , \label{eq-98} \en with the quantity (of order unity)
$\eta$ given by Eq. \ref{eq-gtilde} in Appendix C. Thus, \be U'_m
= \frac{f \, \eta}{2 \, \pi \, c} \, \left( \frac{r_m}{R_m}
\right)^2 \frac{L'_S \, \Gamma^4}{d^2} \, \Gamma^2 .
\label{eq-99}\en}
%
For a moving mirror the calculation of $U'$  proceeds through
steps analogous to the static  case, with proper modifications.}
 In the laboratory frame, the energy density of the radiation
emitted forward at time $t_1$ by the plasmoid at height $z_1$ is,
$U(r_*) = L'_S \, \Gamma^2 / (\pi \, \theta^2 \, c \,
 r_*^2)$,
where $L'_S$ is the initial luminosity in the plasmoid co-moving
frame, and $r_*$ is the distance at which the mirror reflection
occurs, i.e., in the formalism of the Appendix A, $r_* = Z(t_r) -
z_1 $, that becomes
\be r_*  = [\Delta z \, (1 + \beta_o)
  + \beta_o \, c \, (t_1 - t_o)] / (1 - \beta_o) .  \label{eq-r*}\en
We assumed that in the observer's frame the plasmoid emission is
beamed  {into} a forward cone of semi-aperture $\theta $. At time
$t_r$ (see Eq. \ref{eq-tr}  in  Appendix A), a fraction $f$ of the
impinging luminosity is reflected back by the moving mirror, with
luminosity $L_m$ given by $ L_m = \pi \, r_m^2 \, c \, f \, U(r_*)
$, where $r_m$ is the plasmoid size, and $f \sim 0.1$. We then
have \be L_m = \pi \, r_m^2 \, c \, f \, \frac{L'_S \,
\Gamma^2}{\pi \, \theta^2 \, c \, r_*^2 } \simeq \frac{f \,
L'_S}{4} \, \left( \frac{r_m}{D_o} \right)^2 \, \left(
\frac{\Gamma}{\Gamma_o} \right)^4 \label{eq-lm} ,\en where we used
the relativistic beaming relation $\theta \sim 1/\Gamma$, and from
Eq. \ref{eq-r*} with $\Delta z = 0$ and $D_o = \beta_o \, c \,
(t_1 - t_o) $, \be r_* = 2 \, D_o \, \Gamma_o^2 . \en Eq.
\ref{eq-lm} can therefore be used to calculate the photon energy
density of the reflected radiation at the "backward" distance
$\tilde{d}$, $U_m$, that in the laboratory frame for isotropic
emission in the backward half-hemisphere is \be U_m = \frac{L_m}{2
\, \pi \, \tilde{d}^2 \, c} . \label{eq-14} \en

Finally, the reflected photon energy density in the plasmoid
comoving frame at the moment of Compton up-scattering $t_C$ reads
{\mtbb (see also Appendix C and Eq. \ref{eq-gtilde} properly
modified for a moving mirror)}
\be U'_m =  \tilde{\eta} \, U_m \, \Gamma_r^2 =
 \frac{f \, \tilde{\eta} \, L'_S}{2 \, \pi \, c} \, \left( \frac{r_m}{D_o}
\right)^2 \, \frac{1}{4} \left( \frac{\Gamma}{\Gamma_o} \right)^4
\, \frac{1}{\tilde{d}^2}  \, \Gamma_r^2 ,\en {\mtbb where
$\Gamma_r$ is the \emph{relative} Lorentz factor  between the
moving mirror and plasmoid given by Eq. \ref{eq-gamrel} of
Appendix C. For values of $\Gamma_o^2 \gg 1$ and $\Gamma^2 \gg 1$
{\mtv with $\Gamma > \Gamma_o$}, we have $\Gamma_r \simeq \Gamma/
2\, \Gamma_o$ (see Appendix C). Therefore,  \be U'_m = \frac{2 \,
f \, \tilde{\eta} }{\pi \, c} \, \left( \frac{r_m}{D_o} \right)^2
\, \frac{ L'_S \, \Gamma_r^4}{\tilde{d}^2}  \; \Gamma_r^2
\label{eq-110} .\en {\mtbb Note that this last equation for $U'_m$
for the moving mirror case is similar to the static mirror (Eq.
\ref{eq-99}) except for the kinematic factor $4 \, \tilde{\eta}$,
{\mtyy the dependence of $d$ and $\tilde{d}$ on $\Gamma$,} {\mtv
and the obvious substitution $\Gamma \rightarrow \Gamma_r$}.}
{\mtyy On using} Eq. \ref{eq-tilde3} {\ff and $\Delta \, z = 0$},
we obtain
 \be U'_m  = 
 \frac{f \, \tilde{\eta} \, L'_S}{2 \, \pi \, c} \, \left( \frac{r_m}{D_o}
\right)^2 \, \frac{1}{D_o^2}  \, \left( \frac{\Gamma}{\Gamma_o}
\right)^8  \, \Gamma_r^2 \, \label{eq-100} . \en  In terms of
$\Gamma_r$ we have \be U'_m =
 \frac{ 2^8 \, f \, \tilde{\eta}}{2 \, \pi \, c} \, \left( \frac{r_m}{D_o}
\right)^2 \, \frac{L'_S \, \Gamma_r^4}{D_o^2}  \; \Gamma_r^6
\label{eq-101}. \en }
 {\mtdd Note that powers of $\Gamma$ in Eqs.
\ref{eq-100} and \ref{eq-101}  appear to be different from those
of Eq. \ref{eq-110}, but in fact, these equations  are consistent
once the appropriate definition of the distance $\tilde{d}$ is
used (see Eq. \ref{eq-tilde2}).}

 {\mtbb For  values that can be reasonably
applied to our situation, $r_m \simeq 3 \cdot 10^{16} \, $cm, $D_o
\simeq 3 \, r_m$, $f = 0.1 \, f_{-1}$, and $\Gamma_o = 3$}, we
have \be U'_m \simeq (f_{-1} \, 10^{-7} {\rm \; erg \, cm^{-2}})
\, L'_{S,42} \, \frac{1}{ (\Gamma_o/3)^2} \, \left(
\frac{\Gamma}{\Gamma_o} \right)^8 \, \Gamma^2 \label{eq-u'}, \en
where $L'_{42} = L'_S /(10^{42} \rm \, erg \, s^{-1})$.

 It is interesting to compare Eq. \ref{eq-u'} with a typical
photon energy density in the plasmoid comoving frame inside the
BLR, \be U'_{BLR} =
\frac{4}{3} \, \frac{\xi_{BLR} \, L_D \, \Gamma^2}{4 \, \pi \,
R_{BLR}^2 \, c} \simeq (0.03 \; {\rm erg \, cm^{-3}}) \, L_{D,46}
\, \xi_{BLR,-1} \, \Gamma^2 , \en where the disk luminosity is
$L_{D,46} = L_D /(10^{46} \rm \, erg \, s^{-1})$,  $R_{BLR} = 3
\cdot 10^{17} \rm \, cm$, {mtdd and $\xi_{BLR} = 0.1 \,
\xi_{BLR,-1}$ is the BLR average covering factor}. Note that
 $U'_m > U'_{BLR} $ for \be \frac{\Gamma}{\Gamma_o} \gtrsim 4 .
\en  Therefore, a plausible configuration for the moving
mirror-plasmoid geometry 
{\ff is} obtained by the following parameters: $ \Gamma \simeq 10
- 15, \Gamma_o \simeq 2-3, r_m \simeq 3 \cdot 10^{16} \, \rm cm$.

\section{Appendix: Angular Factors}

In this Appendix we calculate the angular factor $\eta$ of the
photon energy density $U'$ in the plasmoid comoving frame that is
dependent on the mirror-plasmoid distance $d$. Unprimed quantities
are calculated in the laboratory frame; primed quantities refer to
the plasmoid comoving frame.

Let us {\mtv first obtain} 
 angular factor $\eta$ for a static mirror. Denoting by $U_m =
L_m/2 \, \pi \, c \, d^2$ the photon energy density at  a distance
$d$ generated by the mirror reflection, we calculate  $U'$
{\ff by integrating over the specific energy density $ U_m/2
\pi$,}
 \be U' =  U_m \, \int_{\mu'_1}^{\mu'_2} \frac{
d\mu'}{\Gamma^4 (1 + \beta\mu')^4} =  U_m \, \int_{\mu_1}^{\mu_2}
{\Gamma^2 (1 - \beta\mu)^2} \, d\mu  \en where the energy Lorentz
transformation reads $\epsilon = \Gamma \, \epsilon \, (1 +
\beta\mu')$, $\epsilon' = \Gamma \, \epsilon (1 - \beta \mu)$, and
$\mu$ and $\mu'$ are the direction cosines in the laboratory and
comoving frames, respectively (related by $\mu' = (\mu - \beta)/(1
- \beta\mu)$). In our case \be U' =  U_m \, \int_{-1}^{-\mu^*}
{\Gamma^2 (1 - \beta\mu)^2} \, d\mu =  U_m \, \Gamma^2 \, \left[ -
\, \frac{1}{3 \, \beta} \, (1 - \beta\mu)^3 \right]_{-1}^{-\mu^*}
\en with $\mu^*$ the upper extreme of integration that depends on
the relative sizes of $r_m$ vs. $d$, i.e.,  \be \mu^* =
\frac{d}{\sqrt{r_m^2 + d^2 }} . \en By defining $\beta_d = \beta
\, \mu^*$, we have \be U' =  U_m \, \frac{1}{3 \, \beta} \,
\Gamma^2 \, \left[ (1 + \beta)^3 - (1 + \beta_d)^3 \right] .\en We
can then define the function $\eta$ that takes into account the
angular integration in the calculation of  $U'$, \be U' = \eta \,
U_m \, \Gamma^2 \,
 ,\en with \be \eta = \frac{1}{\beta} \, \left[
\beta - \beta_d + \beta^2 - \beta_d^2 + \frac{1}{3} \, (\beta^3 -
\beta_d^3) \right] , \en which can be rewritten as \be \eta =
1 - \mu_* + \beta (1 - \mu_*^2) + \frac{1}{3} \, \beta^2
\, ( 1 - \mu_*^3) 
. \label{eq-gtilde} \en

For a moving mirror, the calculation of the angular factor
proceeds in a similar way, with the replacements in the above
expressions:  of $\eta \rightarrow \tilde{\eta}$    $d \rightarrow
\tilde{d}$, $\beta \rightarrow \beta_r$, and $\Gamma \rightarrow
\Gamma_r$;  $\beta_r$ and $\Gamma_r$ are related by the usual
relation $1 - \beta_r^2 = 1/\Gamma_r^2$. The quantity $\Gamma_r$
is the relative bulk Lorentz factor between mirror and plasmoid
and is obtained by the Lorentz transformation in the mirror frame
\be \Gamma_r = \Gamma \, \Gamma_o \, (1 - \beta \, \beta_o) .
\label{eq-gamrel} \en It is useful to note that for $\Gamma^2 \gg
1$, $\Gamma_o^2 \gg 1$, we can approximate the relative Lorentz
factor as \be \Gamma_r \simeq \frac{1}{2} \left(
\frac{\Gamma}{\Gamma_o} + \frac{\Gamma_o}{\Gamma} \right) . \en
{\mtyy For $\Gamma > \Gamma_o$, {\mtxy the simple} approximation
$\Gamma_r \simeq \Gamma / 2 \, \Gamma_o$ can be used.}

\newpage

\end{document}